\documentclass[12pt,a4paper]{article}
\usepackage{geometry}                % See geometry.pdf to learn the layout options. There are lots.
\usepackage{graphicx}
\usepackage{amssymb,amsmath}
\usepackage{epstopdf}
\usepackage{marvosym}
\usepackage[margin=30pt,font=small,labelfont=bf]{caption}
\usepackage[titles]{tocloft}
\renewcommand{\baselinestretch}{1.4}
\def\be{\begin{eqnarray}}
\def\ee{\end{eqnarray}}
\def\dd{{\rm d}}

\def\cD{{\cal D}}
\def\cL{{\cal L}}

\newcommand{\eqn}[1]{(\ref{#1})}

\begin{document}
%\maketitle
\begin{titlepage}

\begin{flushright}
MIT-CTP 4019\\
Imperial/TP/2009/JS/01
\end{flushright}
\vfil

\begin{center}
{\huge A Rotating Holographic Superconductor}\\

\end{center}
\vfil
\begin{center}
{\large \textsc{Julian Sonner}}\\
\vspace{1mm}
{\it Blackett Laboratory, Imperial College, London SW7 2AZ, UK }\\
{\it and}\\
{\it  Trinity College, University of Cambridge, Cambridge CB2 1TQ, UK} \\
{\tt js499@cam.ac.uk}\\
\vspace{3mm}
\end{center}

\vfil

\begin{center}

{\large \textbf{Abstract}}
\end{center}

\noindent
In this paper we initiate the study of SSB in $3+1$ dimensional rotating, charged, asymptotically AdS black holes. The theory living on their boundary,  $\mathbb{R}\times S^2$, has the interpretation of a $2+1$ dimensional rotating holographic superconductor. We study the appearance of a marginal mode of the condensate as the temperature is decreased. We find that the transition temperature depends on the rotation. At temperatures just below $T_c$, the transition temperature at zero rotation, there exists a critical value of the rotation, which destroys the superconducting order.  This behaviour is analogous to the emergence of a critical applied magnetic field and we show that the superconductor in fact produces the expected London field in the planar limit.
\vfill
\begin{flushleft}
March 2009
\end{flushleft}
\vfil
\end{titlepage}
\newpage
\renewcommand{\baselinestretch}{1.1}  %looks better

%%%%%%%%%%%%%%%%%%%%%%%%%%%%%%%%%%%%%%%%%%%%%
%% include the next line for double spacing %%
%%%%%%%%%%%%%%%%%%%%%%%%%%%%%%%%%%%%%%%%%%%%%%
%\renewcommand{\baselinestretch}{2}
\renewcommand{\arraystretch}{1.5}

\section{Introduction}
Over the past year or so, there has been considerable activity in the field of AdS/CMT, which attempts to bring a phenomenological approach to the AdS/CFT correspondence to bear on problems in condensed-matter physics \cite{Hartnoll:2007ip,Hartnoll:2007ih,Balasubramanian:2008dm,Son:2008ye,Adams:2008wt,KeskiVakkuri:2008eb,Gubser:2008wz,Gubser:2008wv}. The scope of this program is broad, dealing with such diverse problems as the Nernst effect at a superfluid-insulator transition, the quantum Hall effect and cold atoms in the unitarity limit. 

In \cite{Gubser:2008px}, Gubser suggested that spontaneous $U(1)$ symmetry breaking by bulk black holes - specifically Reissner-Nordstr\"om-AdS black holes -  can be used to construct gravitational duals of the transition from normal state to superconducting state in the (not further specificed) boundary theory. This can be done by studying the dynamics of a minimally coupled complex scalar field with Lagrangian density
\be
\frac{L}{\sqrt{-g}} =  -\frac{1}{4}F^2- \frac{1}{2}\left|{\cal D}\Psi \right|^2 - m_\Psi^2 \left|\Psi\right|^2 \,,
\ee
where ${\cal D}_\mu\Psi = (\partial_\mu - ie A_\mu)\Psi$ and the scalar field mass respects the Breitenlohner-Freedman (BF) bound \cite{Breitenlohner:1982jf}
\be
m_\Psi^2 L^2 \geq -9/4\,.
\ee
In the remainder of this paper, we only consider scalar masses at or above the conformal value of $m^2_\Psi=-2$.
In the first instance one works in the approximation where the scalar does not backreact on the geometry. This approximation to the full system is however sufficient for an {\it exact} determination of the transition temperature of the holographic superconductor.

Gubser showed numerically that for sufficiently large charge $e$ of the condensate the scalar field on the black hole background allows a marginal mode. This is indicative of an instability towards a `hairy' black hole solution with a finite charged condensate outside the horizon. The scalar hair breaks the $U(1)$ gauge symmetry in the bulk space time. Hartnoll, Herzog and Horowitz  \cite{Hartnoll:2008vx} brought this idea to fruition by numerically constructing the state with broken symmetry. They were able to show that the electrical resistance in the dual field theory indeed drops to zero in the broken phase by studying the fluctuations of the Maxwell field around the state of broken symmetry.

In subsequent works, these authors, and others  \cite{Hartnoll:2008vx,Albash:2008eh}, verified a number of other physical properties of these backgrounds and accumulated evidence that these are gravitationals dual of a type II superconductor. This is despite the fact that a {\it local} bulk $U(1)$ symmetry is a priori associated to a {\it global} symmetry of the boundary theory, and hence one would expect to find the physics of a charged superfluid \cite{Herzog:2008he}. However one may weakly gauge the boundary $U(1)$ and in this regime the material is described by the London theory \cite{Hartnoll:2008vx} of superconductivity.

 In this paper we aim to put to explore another feature of superconductors, the {\it London moment}, and by doing so, extend the present framework of holographic superconductors to the case of rotating black holes in the bulk. At first sight, this also seems to be a hopeless task, since the holographic superconductor does not have a dynamical photon. However, the work of  \cite{Hartnoll:2008vx} showed that it nevertheless produces the screening currents necessary {\it e.g.} for the Meissner effect. Thus we may also be optimistic about being able to observe the London moment.
 
  At zero angular momentum, the system studied here reduces to a subset of those considered in \cite{Gubser:2008px}, with spherical horizon, where the transition to a superconducting state is already known to occur. We thus expect the marginal mode to exist for small values of the rotation parameter. An interesting question is whether this mode always persists for larger values. Physical intuition tells us that it should not: we know from previous studies \cite{Albash:2008eh} that a holographic superconductor exhibits a critical magnetic field, above which it is energetically favourable to be in the normal phase. We also  know that a rotating superconductor dynamically generates a magnetic field, the so-called {\it London field} \cite{London}. Intuitively this arises because the rotation induces a lag between charge carriers close to the surface and the charged superfluid in the bulk of the superconductor. Thus we should expect a given marginal mode at zero rotation to disappear at some critical rotation. Or, in other words, we expect the critical temperature to drop as the angular momentum of the dual black hole is raised. We shall demonstrate this behaviour.

The paper is organised as follows. In section 2 we introduce the dual gravity background, the Kerr-Newman-AdS black hole and explain the influence of the bulk rotation on the boundary theory. In section 3 we study the angular and radial behaviour of marginal modes of a charged complex scalar field in this background and construct the phase diagram of the holographic superconductor. For a tachyonic scalar in the planar limit, {\it i.e.} close to the poles, we compute the London field explicitly. Section 4 is devoted to a discussion and appendices A and B contain details on the separation of the wave equation as well as the numerical methods employed in this paper.

\section{Rotation and AdS/CMT}\setcounter{equation}{0}
In this section we introduce the gravitational background, which we argue is dual to a two-dimensional material living on the surface of a sphere, exhibiting superconductivity at sufficiently low temperature. The study of rotation in the AdS/CFT context was initiated in \cite{Hawking:1998kw} (see also \cite{Berman:1999mh}).
\subsection{AdS$_4$ backgrounds with Angular Momentum}
The desire to introduce rotation into the AdS/CMT story leads us to consider the four-dimensional Kerr-Newman-AdS solution, orginally found by Carter \cite{Carter:1968ks}, who was interested in finding space times in which the Hamilton-Jacobi equation would separate. In Boyer-Lindquist coordinates, the metric is
\begin{equation}
\dd s^2= - \frac{\Delta_r}{\rho^2}\left[ \dd t - \frac{a}{\Xi}\sin^2\theta \dd \phi \right]^2 + \frac{\rho^2\dd r^2}{\Delta_r} + \frac{\rho^2 \dd \theta^2}{\Delta_\theta} + \frac{\Delta_\theta\sin^2\theta}{\rho^2}\left[ a\dd t - \frac{r^2 + a^2}{\Xi}\dd \phi \right]^2
\end{equation}
with gauge field field
\be
A=-\frac{q_er}{\rho^2}\left(\dd t-\frac{a\sin^2\theta}{\Xi}\dd\phi  \right)\,,
\ee
and
\be
\Delta_r := &\bigl(r^2 + a^2\bigr)\bigl(1 + r^2 L^{-2}  \bigr) - 2 M r + q_e^2 \,,\qquad  \Delta_\theta &:= 1 - a^2 L^{-2}\cos^2\theta\nonumber\\
\rho^2 := &r^2 + a^2\cos^2\theta\,\qquad\qquad\qquad\qquad \qquad\qquad  \Xi &:=1 - a^2 L^{-2}
\ee
There is some confusion in the literature as to how exactly the parameters $a$ and $M$ are related to the physical angular momentum and mass-energy of the black hole, respectively. This confusion is cleared up in the work \cite{Gibbons:2004ai} (see also \cite{Papadimitriou:2005ii}), by careful thermodynamic considerations, and we adopt their definitions
\be
E=\frac{M}{\Xi^2}\,,\qquad\qquad J=\frac{Ma}{\Xi^2}
\ee
for the energy $E$ and angular momentum $J$ of the background. Note that both $E$ and $J$ diverge as $aL^{-1}$ approaches unity. However, $J$ is always strictly bounded above by $EL$, as expected for a rotating black hole. Finally, we quote the result for the Hawking temperature of the black hole \cite{Caldarelli:1999xj},
\be\label{eq:thawking}
T_H = \frac{r_+(1 + a^2 L^{-2} + 3 r_+^2 L^{-2} - (a^2 +q_e^2)r_+^{-2})}{4\pi (a^2 + r_+^2)}\,,
\ee
which, in the usual way, is identified with the temperature of the dual field theory living on the conformal boundary of the AdS black hole spacetime. The boundary has topology $\mathbb{R}\times S^2$, so we are considering a two-dimensional superconductor living on the surface of a sphere. The quantity $r_+$ is the horizon radius, defined to be the largest real root of the equation $\Delta_r=0$.
\subsection{Rotation and the Boundary Theory}
Massive rotating bodies exhibit the {\it frame-dragging effect}: an inertial reference frame outside the body is set into rotational motion. This effect diminishes as the distance to the rotating body increases. For the original Kerr black hole, which is asymptotically flat, this effect vanishes at infinity. However, for the Kerr-AdS family, it does not \cite{Gibbons:2004ai}. Its boundary metric is in the conformal class of the Einstein Static Universe
\be
\dd s^2 = -\dd t^2 + L^2 \left( \dd \hat\theta^2 + \sin^2\hat\theta \dd \hat\phi^2 \right)
\ee
where the new angular coordinates satisfy
$$\phi = \hat\phi - \frac{a}{L^2}t\,, \qquad\tan\theta = \sqrt{\Xi}\tan\hat\theta\,.$$ 
From the first of these see that the angular momentum of the boundary theory is
\be
\Omega_\infty = \frac{a}{L^2}\,.
\ee
We learn that the local speed of rotation at the {\it equator} of the boundary $S^2$ of radius $L$ reaches the speed of light when $a=L$. In fact the gravitational background becomes singular at that point. We therefore restrict to $a<L$ in this paper.

The meaning of rotation in the field theory becomes apparent if we compute for example the free partition function, {\it i.e.} the free energy of the theory at zero coupling. It is given by the expression
\be
F=-\ln{\cal Z}=T_{\rm bdry} \sum_i \sum_{\ell=0}^\infty \sum_{m=-\ell}^\ell \eta_i \ln \left(1-\eta_i e^{-\beta\left( \omega - m \Omega_\infty \right)}  \right)\,,
\ee
where $i$ labels the different (particle) species, $\eta_i=\pm 1$ for bosons (fermions) and $\beta=1/T_{\rm bdry}$. The allowed frequencies $\omega$ follow from the corresponding free wave equations. The summation is over the angular momentum quantum numbers of the fields $(\ell,m)$ as indicated. The quantity $\Omega_\infty$ enters as a chemical potential constraining the angular momentum.
\section{Superconducting Instablity}\setcounter{equation}{0}
We now turn to the main objective of this work, which is to identify an instability under perturbations of a charged scalar field propagating on the background. We rely on the separation properties of this equation, but relegate the details of the separation procedure to appendix A. It is not necessary to follow these in order to understand this paper, but we have included them for the interested reader.
\subsection{Radial Equation}
As is well known from a classical result due to Carter \cite{Carter:1968ks}, the Klein-Gordon equation is separable on the background of the Kerr-AdS black hole in $3+1$ dimensions. This result can be extended to the case of a complex (charged) scalar field on the Kerr-Newman-AdS background above. We assume that $\Psi(t,r,\theta,\phi)\sim e^{-i\omega t - i m\phi}R(r) S(\theta)$, and look for a marginal mode, for which $m=\omega=0$.  As in \cite{Gubser:2008px,Hartnoll:2008vx}, it is consistent to seek real solutions $R(r)$ and $S(\theta)$. However, since the complex scalar $\Psi$ is not a gauge-invariant quantity, this is essentially a choice of gauge, the details of which are explained in appendix A. For later convenience, let us define the {\it horizon function}
\be
r^2h(r)=\Delta_r(r)\,.
\ee
Then we obtain the radial equation
\be\label{eq:radialI}
\left(\partial_r r^2h(r) \partial_r   + \frac{ e^2 q_e^2 }{h(r)}\left( 1-r \right)^2 - (a^2+r^2) m_\Psi^2 - \lambda \right)R(r) &=& 0 \,,
\ee
where we have set the horizon radius to unity\footnote{This can always be done because of scaling symmetry present in the background. However, physically, this is nothing but a choice of units in which we measure lengths in multiples of the horizon radius.}, and $\lambda$ is the separation constant between $S(\theta)$ and $R(r)$. In the Schwarzschild-AdS limit ($a\rightarrow 0$) it reduces to the familiar value $\ell(\ell+1)$, $\ell\in\mathbb{Z}$ being the principal eigenvalue of the spherical harmonics. Sometimes we shall find it useful to label the values of $\lambda$ by the corresponding integer $\ell$ to which they reduce. While there exists no analytical expression for $\lambda$, it is possible to determine the numerical value of this separation constant by treating the $\theta$-equation as an eigenvalue problem for $\lambda$.

To gain a more intuitive understanding of the physics involved in finding the marginal mode, let us manipulate  expression \eqn{eq:radialI} into the form of a Schr\"odinger equation. Then the question of finding the gravitational mode is re-expressed in terms of finding a marginal bound state of the corresponding quantum-mechanical problem.
To this end, define the {\it tortoise coordinate}
\be\label{eq:tortoise}
r_* = \int\frac{\dd r}{h(r)}\,.
\ee
%
%Explicitly:
%%
%\be
%r^* = \sum_{r_i=1}^4\frac{1}{\left(h^\Delta(r_i)\right)'}\ln\left( r-r_i\right)\,,
%\ee
%
Outside the horizon at $r=1$, $r_* $ is real and takes values\footnote{The upper bound can be seen from the neat identity $\sum_{\rm roots}\frac{1}{{\cal Q}'(x_i)}=0$ for any polynomial ${\cal Q}(x)$ at least quadratic in $x$. A simple proof \cite{zsuzsa08} can be given via contour integration of the function $f(z)=\frac{1}{{\cal Q}(z)}$ defined by analytically continuing ${\cal Q}$ into the complex plane.} $r_* \in (-\infty,0)$. Then the function $Z(r) = r R(r)$ satisfies the Schr\"odinger equation of a zero-energy particle in a potential:
\be\label{eq:schrod}
\frac{d^2 Z}{dr_*^2} - V\left[r_*(r),\lambda\right] \, Z(r_*)=0\,.
\ee
The potential is given implicitly by
\be\label{eq:potential}
V(r,\lambda) = h (r)\left[ m_\Psi^2 - \frac{e^2 q_e^2}{h(r)}\left( \frac{1}{r}-1 \right)^2  + \frac{\left( h(r) \right)'}{r} + \frac{\lambda + a^2m_\Psi^2}{r^2}  \right]\,,
\ee
where the prime indicates differentiation with respect to the original radial coordinate $r$.
Since we cannot invert $r_*(r)$ analytically to get the potential as a function of $r$, we cannot give an explicit functional form of the potential as a function of the tortoise coordinate $r_*$. It is however easy to obtain plots of the potential, numerically or otherwise, which are sufficient to develop the intuition we wish to achieve. Before proceeding to obtain these plots, we must first return to the problem of the separation constant and the angular equation. In the process we will gain insight into the nature of the condensate.
\subsection{Angular Equation and Localization}
Without further ado, here is the angular equation for our marginal mode:
\be\label{eq:angularI}
\left[\Delta_\theta\frac{\partial_\theta(\sin\theta\sqrt{\Delta_\theta}\partial_\theta)}{\sin\theta\sqrt{\Delta_\theta}} +\sqrt{\Delta_\theta}(\partial_\theta\sqrt{\Delta_\theta})\partial_\theta + a^2 \sin^2\theta m^2_\Psi +\lambda \right]S(\theta) =0
\ee
Again, a detailed derviation  is given in appendix A.
We can gain some understanding of this equation by considering two limits. Firstly, in the non-rotating limit, this equation reduces to the associated Legendre equation, defining the $\theta$-dependent part of the spherical harmonics $Y_{\ell m} (\theta,\phi)$. In this limit, we learn that $\lambda = \ell(\ell + 1)$ and it is a useful check on our computations to note that our results reduce to these integer values in the appropriate limit.

\begin{figure}[ht!]
\begin{center}
\includegraphics[width=0.4\textwidth]{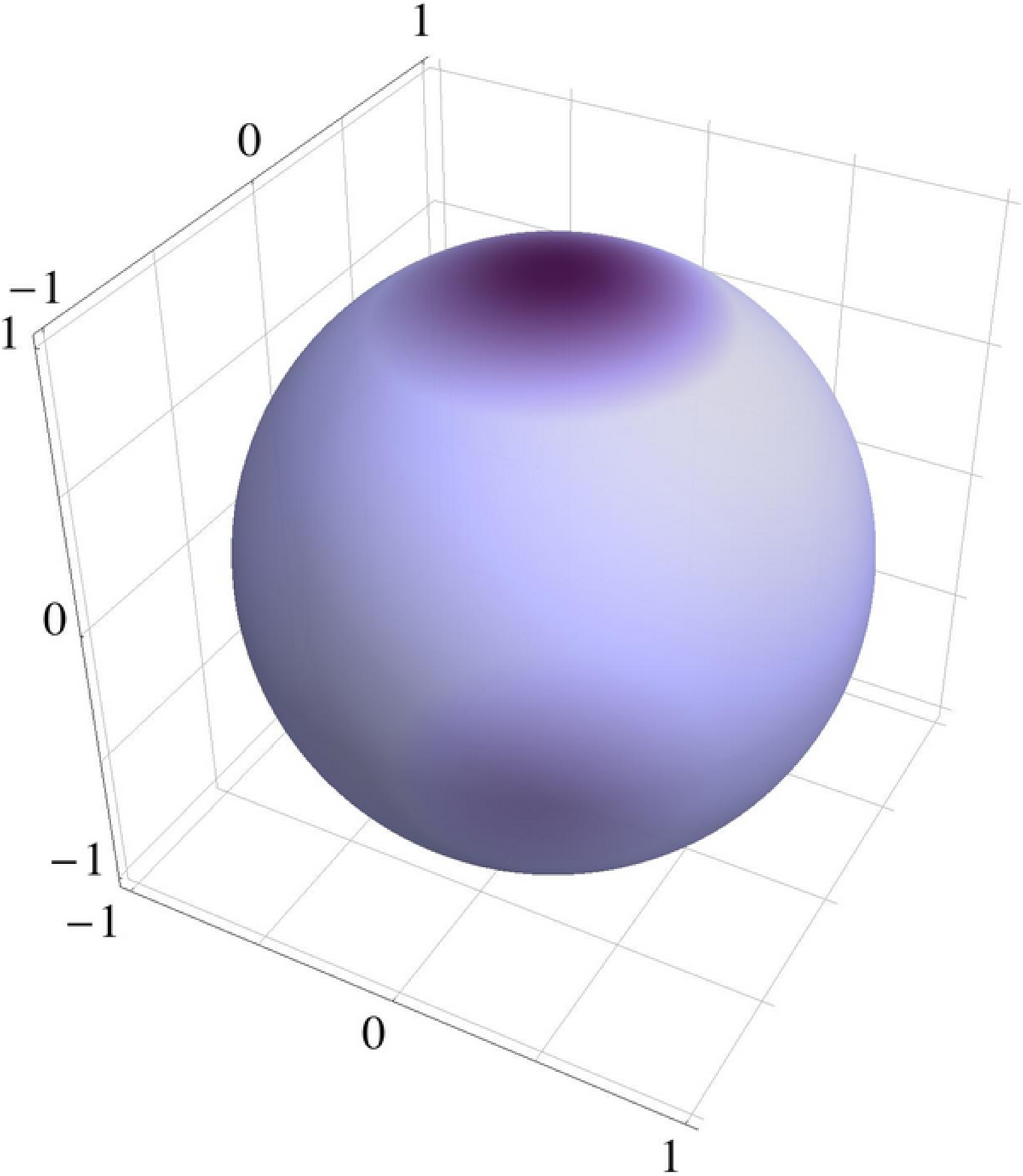}\hfil\includegraphics[width=0.4\textwidth]{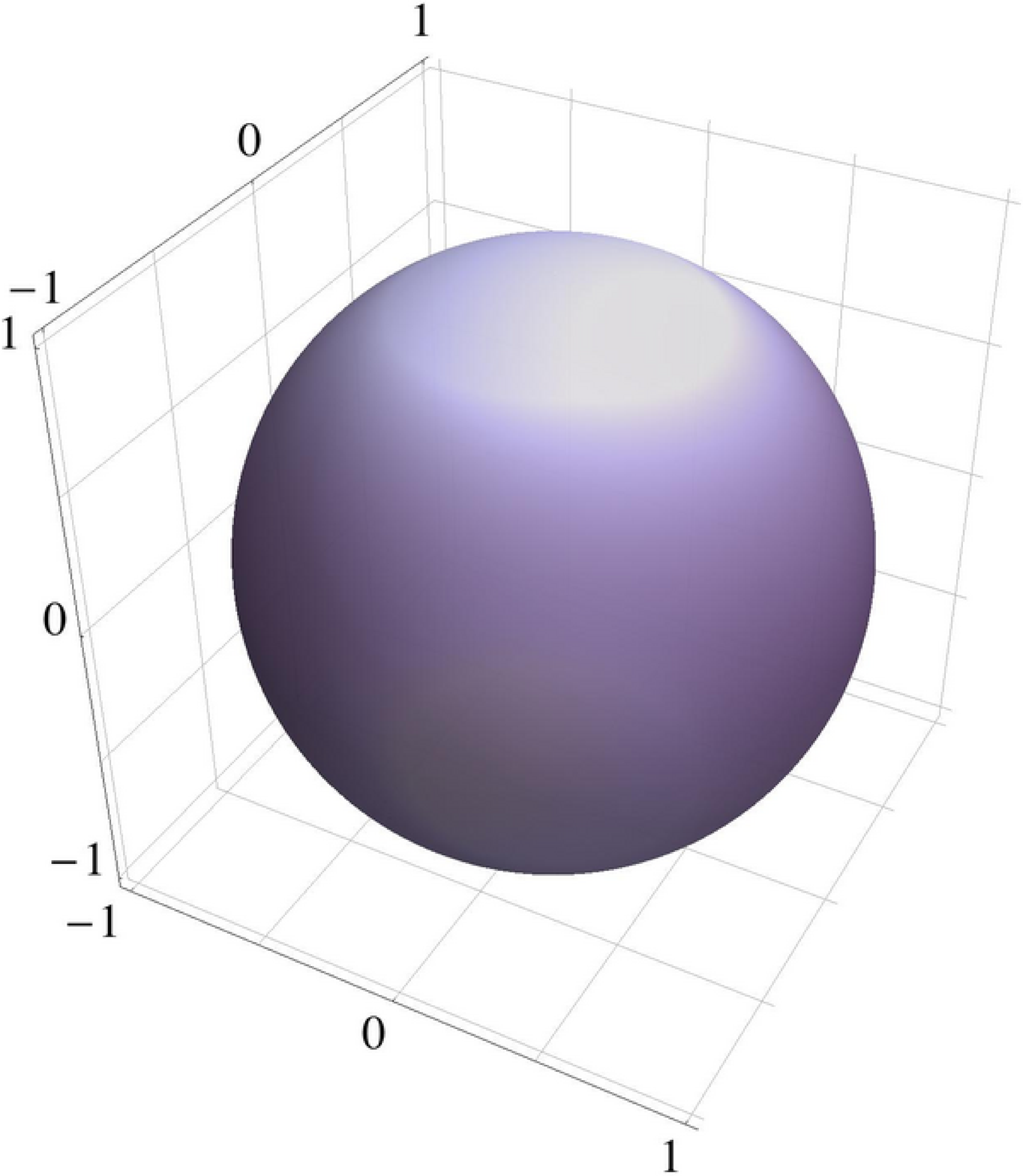}
\caption{{ Lowest angular AdS spheroidal harmonics. Regions with high density of the condensate are shown lighter and regions of low density are shown darker. Both panels correspond to a rotation parameter of $\alpha=0.9$. On the left panel, we have chosen a positive mass-squared term, while on the right the mass-squared term is negative.}}
\label{fig:angular}
\end{center}
\end{figure}

In the flat-space limit, $L\rightarrow \infty$, the angular equation reduces to the {\it spheroidal} equation, first obtained in the black-hole context by Teukolsky \cite{Teukolsky:1972my}. In this limit, no analytical expressions are known for $\lambda$, but its values are tabulated.  Again, it is a confidence-inspiring check on our procedures that our values for $\lambda$ agree with those of the spheroidal harmonics in the appropriate limit. For the sake of terminology, let us refer to the solutions of the angular equation \eqn{eq:angularI} as {\it AdS spheroidal harmonics}. The deviation of the AdS spheroidal harmonics from being spherical harmonics is measured by the ratio of specific angular momentum of the black hole to the AdS length, so we introduce the deformation parameter $\alpha=\frac{a}{L}$.

The eigenvalues of the deformed spheroidal equation \eqn{eq:angularI} are then labelled $\lambda(\ell,\alpha, a^2 m_\Psi^2)$. 
Solutions to equation \eqn{eq:angularI}, much like the associated Legendre functions, have definite parity, which allows us to develop a simple, iterative, shooting technique to solve both for the eigenfunction and the eigenvalue to high accuracy. The details are relegated to appendix B. There we also list a range of eigenvalues $\lambda$, corresponding to various values of the deformation parameter $\alpha$.

An interesting feature of the AdS spheroidal harmonics is that, unlike their French cousins, the Legendre functions, the lowest mode is not constant. Rather it has a non-trivial profile over the sphere, controlled by the deformation parameter $\alpha$ causing even the lowest mode to be localized either near the poles or near the equator. They are localized near the poles if $m^2_\Psi<0$ and near the equator if $m^2_\Psi>0$. This is illustrated in figure \ref{fig:angular}, which shows density plots of the lowest mode on the unit $S^2$. For the special case of $m_\Psi=0$ they are constant.

We now have all the ingredients needed to obtain the critical behaviour of the rotating superconductor. We start by exhibiting marginal modes of the charged scalar field.
\subsection{Marginal Modes}
\subsubsection{Qualitative Considerations}

\begin{figure}[ht!]
\begin{center}
a)\includegraphics[width=0.5\textwidth]{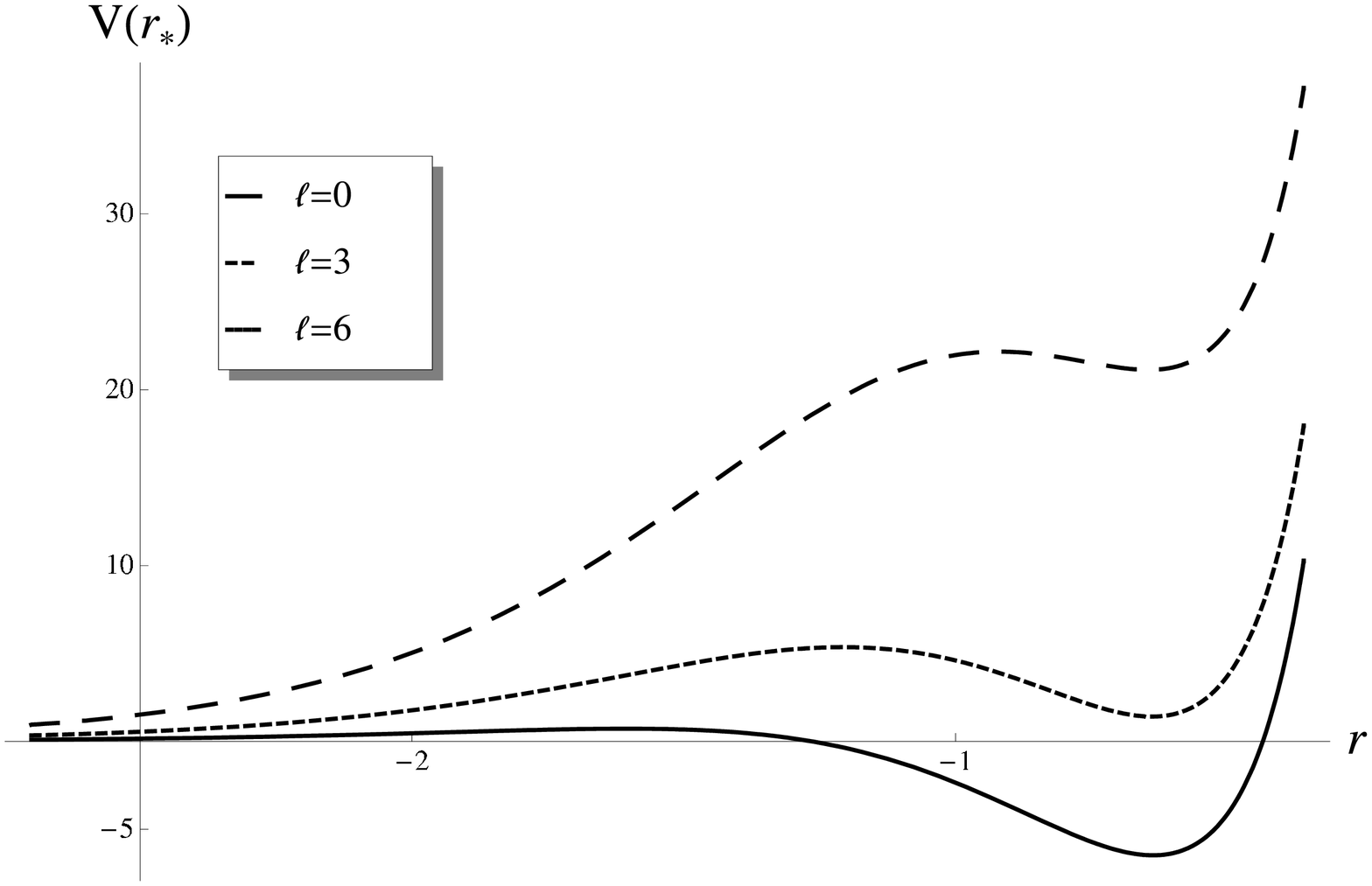}\hfil b)\includegraphics[width=0.3\textwidth]{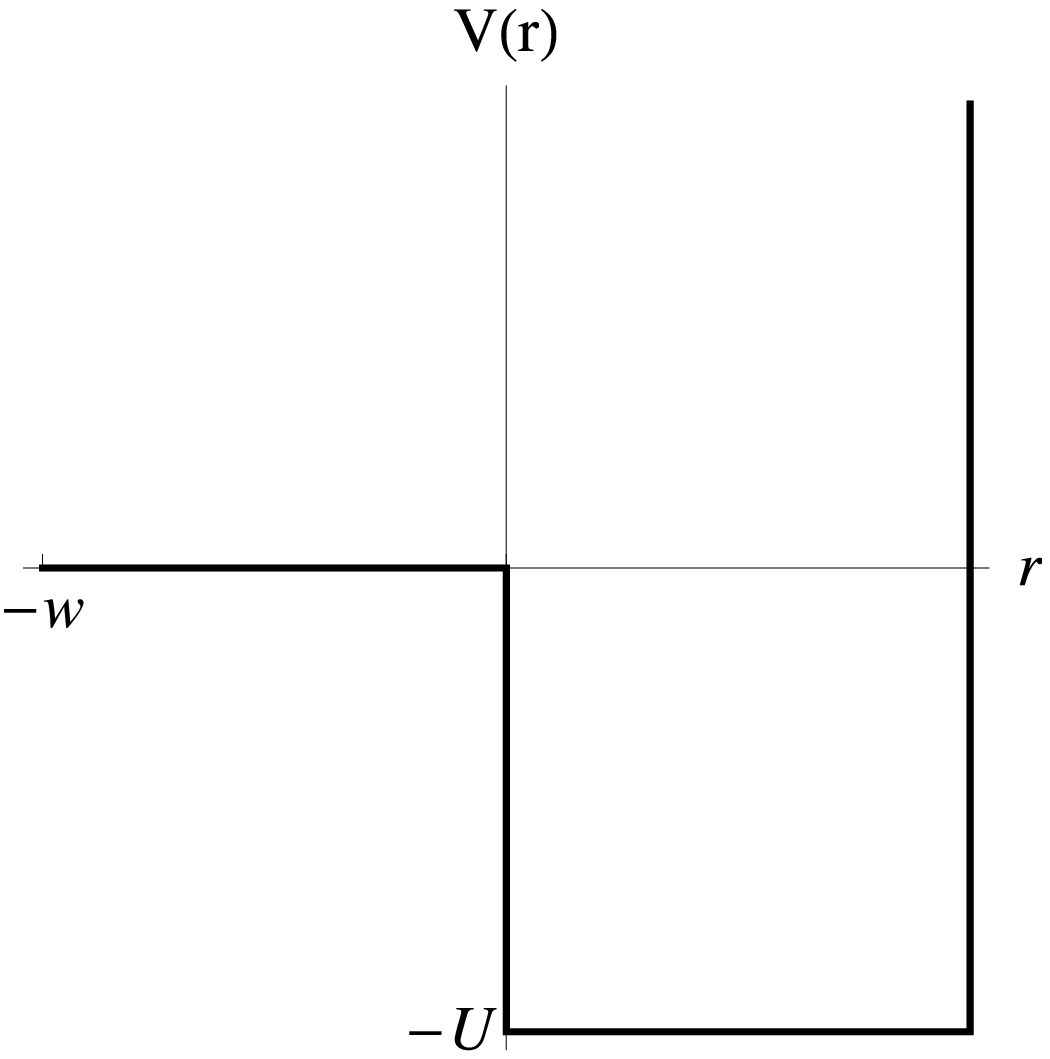}
\caption{a) Plot of effective potential barrier around Kerr-Newman-AdS black hole for charged scalar waves with $m^2_\Psi=4$. In the chosen coordinates, the horizon is at $r_*=-\infty$ and asymptotic infinity is on the right at  $r_*=0$. b) Square well toy model. The depth of the well is $U$ and its width is $w$. AdS boundary conditions are equivalent to putting a reflecting barrier at the origin. }
\label{fig:figure1}
\end{center}
\end{figure}

 Evidently, equation \eqn{eq:schrod} cannot have bound state solutions, unless the potential $V(r_*)$ develops a negative well in a certain range of parameters. Therefore we will expect the lowest marginal mode to lie in the sector with $\ell=0$. Figure \ref{fig:figure1}a illustrates this point. The occurrence of a negative potential well is closely related to the fact, pointed out by Gubser \cite{Gubser:2008px}, that the effective mass of the scalar field
\be
m_{\rm eff}^2:=m_\Psi^2 - \frac{e^2 q_e^2}{h(r)}\left( \frac{1}{r}-1 \right)^2
\ee
 gets a contribution due to the coupling to the background gauge potential that can make this sufficiently negative for a sufficiently long interval outside the horizon. 
 
The criterion for instabilty can be illustrated by considering the toy model of the semi-infinite square well of depth $U$ and width $w$. The system will be stable if the potential admits negative-energy bound-state solutions and will become unstable once the last of these boundstates (i.e. the ground state) acquires positive energy and becomes non-normalizable. Elementary quantum mechanics tells us that the condition that there exists at least one bound state is
$$w\frac{\sqrt{2mU}}{\hbar} \geq \frac{\pi}{2}\,,$$
 where we have briefly reinstated units of $\hbar$. Thus if the potential well is too shallow or too narrow, no bound-state can exist. In particular, for the toy model, a bound state exists for $\nu\leq 4  w$, where $\nu$ is the de-Broglie wavelength of a particle of energy $U$. In searching for the zero-energy marginal bound state, we precisely capture the moment where the ground state exits the Hilbert space of the theory. We interpret this as the transition point for the condensation of the order parameter.

\begin{figure}[t!]
\begin{center}
\includegraphics[width=0.5\textwidth]{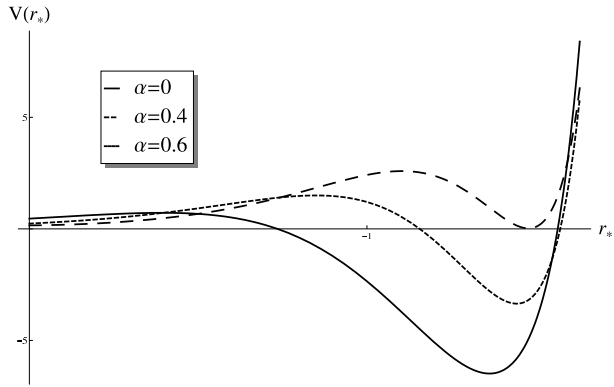}
\caption{{\small Plot of effective potential barrier for the $\lambda$ value corresponding to $\ell=0$ and $eL=10, m^2_\Psi=4$. The potential is given as a function of the deformation parameter $\alpha=aL^{-1}$ at {\it constant} temperature $T_0$. We clearly see that the potential well disappears as one increases the rotation.}}
\label{fig:figure2}
\end{center}
\end{figure}

With this simple picture in mind, let us return to the physics of the holographic superconductor. Keeping the system at at a {\it fixed} temperature $T$ below its transition temperature for zero rotation, $T_0$, the potential well becomes more and more shallow as the specific angular momentum $a$ is increased.
In figure \ref{fig:figure2} this can be seen as the decrease in depth of the potential well, as one increases the rotation $a$ in units of the AdS length. We see that the well becomes both less deep and more narrow as the rotation is increased. Hence the expectation is that the zero-energy mode will cease to exist at some point. 

This section has illustrated the physical mechanism behind the suppression of superconductivity with rotation, but because of the complicated shape of the potential \eqn{eq:potential}, it is necessary to analyze the marginal mode in detail, in order to gain a quantitative understanding of the phenomenon. The issue of stability of AdS black holes under scalar field perturbations is an interesting subject in its own right. For recent mathematical results on (real) scalar fields in AdS black hole backgrounds, see \cite{Holzegel:2009ye}.

\subsubsection{Numerical Solutions}
We will now embark on a study of the bevhaviour of the marginal mode as a function of the parameter $\alpha$. It should come as no surprise to the reader that the only successful  quantitative approach approach to solving the radial equation is the numerical one. Already in the much simpler case of RNAdS black holes, no analytical solutions are known for the marginal mode, much less for the broken phase.

In order to facilitate the numerical analysis, let us transform to the variable $z=r^{-1}$.
Then the question of the marginal mode is transformed into a boundary value problem on the interval $(0,1)$. We wish to solve the equation
\be\label{eq:zradial}
\frac{d}{dz}\left( h(z) \frac{dR}{dz} \right) + \left[\frac{e^2 q_e^2(z-1)^2}{z^4h(z)}  -\frac{m_\Psi^2 (a^2 z^2 +1)}{z^4}-\frac{\lambda}{z^2}\right] R(z)=0
\ee
subject to certain boundary conditions.
We see that both the horizon at $z=1$ and asymptotic infinity at $z=0$ are regular singular points of this equation.
Near the AdS boundary at $z=0$, the solutions take the familiar form
\be
R(z)\sim \Psi_1 z^{\Delta_+} +\Psi_2 z^{\Delta_-}\,,
\ee
where $\Delta_\pm$ is the conformal weight of the dual operator, given by the root of the indicial equation
\be
\Delta(\Delta-3)=m_\Psi^2 L^2\,.
\ee
Unitarity requires that $\Delta\geq\frac{1}{2}$. Depending on the bulk scalar mass, either or both of the operators
\be
\langle {\cal O}_1 \rangle &=& \sqrt{2}\Psi_1 \nonumber\\
\langle {\cal O}_2\rangle &=& \sqrt{2}\Psi_2\,
\ee
may condense. We have chosen the same normalization for the bulk-boundary coupling as \cite{Hartnoll:2008vx}. The indicial equation at the horizon tells us that the solution there behaves as
\be
R(z) \sim R_{(0)} + R_{(1)}\ln (z-1)
\ee
and regularity demands that $R_{(1)}=0.$ 

In order to solve \eqn{eq:zradial}, we set either of $\Psi_i=0$, corresponding to either $\langle {\cal O}_1\rangle$ or $\langle {\cal O}_2 \rangle$ condensing. Thus we develop the solution in a Frobenius series around infinity
\be
R(z)= z^{\Delta}\left( \Psi_{i}+\sum_{n}a_nz^n \right)\,\qquad i=1\text{ or }2
\ee
 This can be done to very high order using a symbolic algebra package to solve recursively for the coefficients.  Similarly the expansion around the horizon takes the form
\be
R(z) = 1 + \sum_{n}b_n(z-1)^n\,,
\ee 
 {\it i.e.} we have chosen to normalize $R_{(0)}=1$. As remarked in \cite{Gubser:2008px} it would be more proper to treat $R_{(0)}$ as a small parameter that allows us to justify neglecting the backreaction, but in the end this is a mere rescaling of the mode.
 
We use numerical integration to match the two series solutions at some intermediate point by adjusting the coefficient $\Psi_i$. This shooting technique can be implemented\footnote{As with all numerical procedures described in this paper, I shall make Mathematica\texttrademark notebooks available upon request.} very efficiently using a numerical root finder and looping over initial conditions.
\begin{figure}[ht!]
\begin{center}
\includegraphics[width=0.4\textwidth]{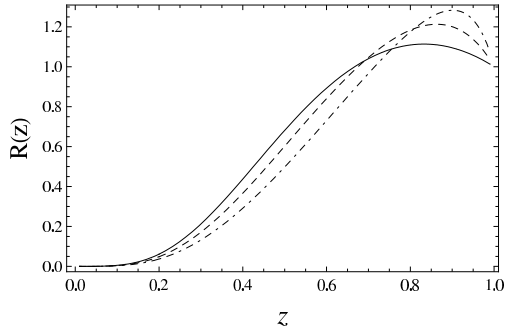}
\caption{Examples of lowest ({\it i.e.} no nodes) marginal modes  for condensation of ${\cal O}_1$. We take $eL=8$, $m^2L^2=4$ and $\alpha=0,0.7,0.9$ for the solid, dashed and dot-dashed modes respectively. In the Schr\"odinger representation these are the radial wavefunctions of the zero-energy bound states.}
\label{fig:marginalmodes}
\end{center}
\end{figure}

However, after matching the value of the function, there is no guarantee that the first derivatives will also match at this point. This is because there is only a marginal mode at certain specific values of the background parameters. Evaluating $T_H$, given in equation \eqn{eq:thawking}, at the critical values for which the mode first appears gives us the critical temperature. An illustration of several marginal modes is given in figure \ref{fig:marginalmodes}.

The family of backgrounds we are interested in depends on the parameters $L,a,q_e, r_+$ and the scalar field has mass $m_\Psi$ and charge $e$. We have chosen units such that $r_+=1$ and analyse the equation for fixed values of $e$ and $m_\Psi$. This results in a curve of critical points in the $(\alpha,q_e)$-plane. For concreteness we fix $q_e=1$ to extract a value for $T_c$. Thus we are looking at projections of the full phase diagram. It would be interesting to extend this analysis to explore the entire phase diagram.

\subsection{Phase Diagram}
In order to find the critical temperature at a given value of background parameters, we follow the example of \cite{Denef:2009tp} and focus on the condensation of ${\cal O}_1$ for different parameter values. This is convenient because it always corresponds to a normalizable solution, as long as $m^2_\Psi \geq -\frac{9}{4}$. Quantitatively, this only gives a lower bound on the critical temperature, since in the mass range where both ${\cal O}_1$ and $ {\cal O}_2$ are normalizable, ${\cal O}_1$ may in fact condense first. However,  in this paper we are not so much concerned with the actual value of $T_c$, but rather its behaviour as a function of $aL^{-1}$.

\begin{figure}[ht!]
\begin{center}
a) \includegraphics[width=0.45\textwidth]{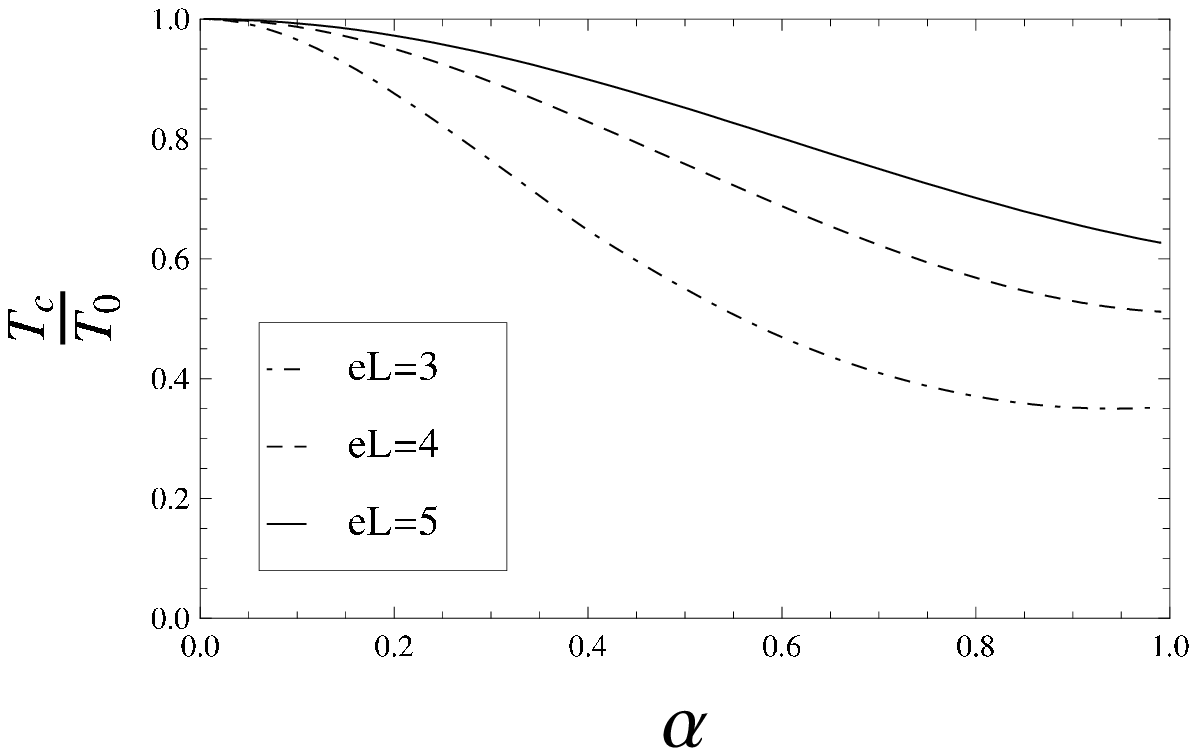}\hfil b)\includegraphics[width=0.45\textwidth]{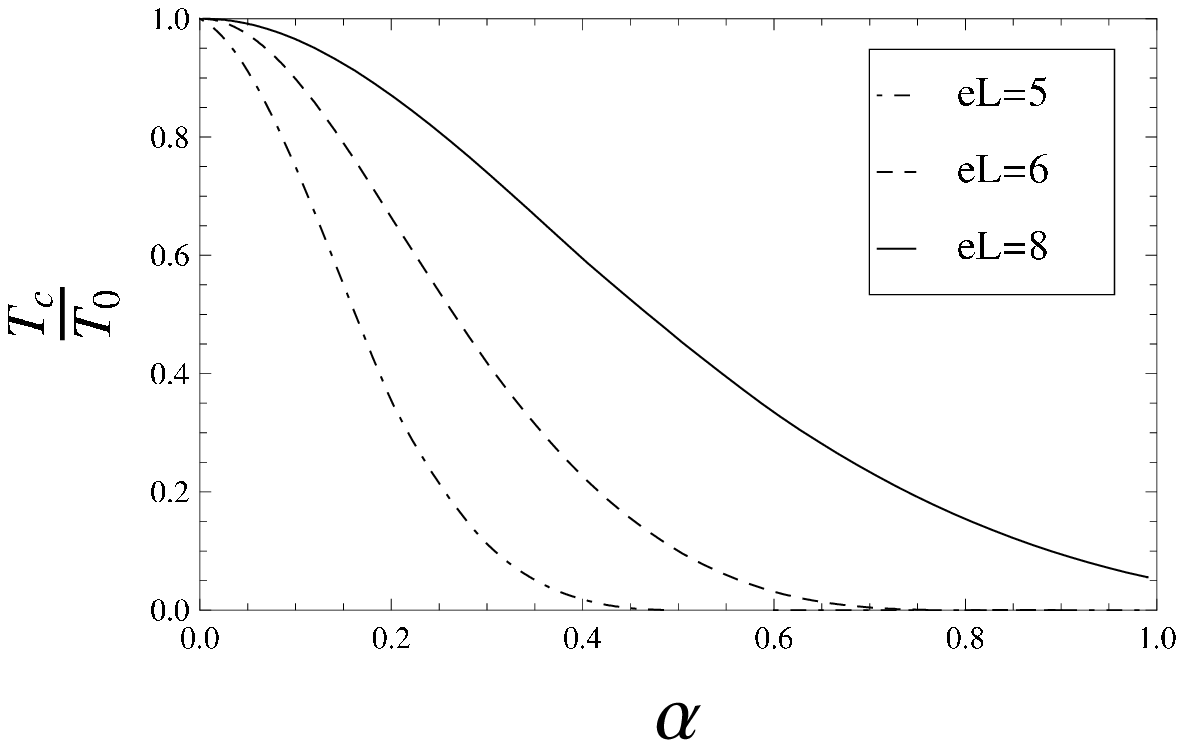}
\caption{{\small Critical temperature for the condensation of ${\cal O}_1 $ at $q_e=1$ as a function of $\alpha=a/L$. Left panel for $m_\Psi^2 L^2=-2$ and right panel for $m_\Psi^2 L^2 = 4$. Superconductivity is increasingly suppressed as the rotation is increased.}}
\label{fig:phasediagram}
\end{center}
\end{figure}

We find that in all cases $T_c$ decreases as the angular momentum of the superconductor is increased. This means that at a temperature slightly below $T_0$, the critical temperature at zero rotation, there always exists a critical value of $aL^{-1}$ above which the material is forced back into its normal phase. 

However, for some choices of parameters, such as in figure \ref{fig:phasediagram}a, one can choose the temperature low enough, that no amount of rotation will force the material back into its normal state. Recall that for $aL^{-1}=1$ the rotation speed at the equator equals the speed of light. However, as we have seen in figure \ref{fig:angular}, the condensate may be localised away from the equator, close to the poles, where the local speed of rotation is arbitrarily small. Thus, it is plausible, that in these cases, enough of the condensate recedes far enough away from the areas of the $S^2$, where the rotation would destroy the superconducting order. 

\subsection{Planar Limit and London Moment}
Let us consider what happens if we concentrate on a small region around the north pole of the boundary $S^2$. To this end, we introduce the coordinate
\be
u = L\hat\theta\,,
\ee
and consider the regime $u\ll L$, with $\alpha=\frac{a}{L}$ held constant. In other words we zoom in on the physics close to the north pole on scales much smaller than the AdS length. In this regime the curvature of the sphere is negligible and we are effectively dealing with a flat horizon rotating with angular velocity $\Omega = \frac{a}{L^2}$. Note that the local speed of rotation is of order ${\cal O}\left( uL^{-1} \right)$, small compared to the speed of light. The angular equation \eqn{eq:angularI} simplifies to give
\be
S''(u) + \frac{1}{u}S'(u) + \frac{a^2 m_\Psi^2u^2}{\Xi L^4} S(u) = -\frac{\lambda}{\Xi L^2}S(u)\,.
\ee
Compare this with the equation of a planar holographic superconductor immersed in a magnetic field, viz.
\be
S''(u) + \frac{1}{u}S'(u) - \left(\frac{euB}{2}\right)^2S(u) = -\tilde\lambda S(u)\,,
\ee
for a different separation constant of mass dimension two, as defined in \cite{Hartnoll:2008vx}. This equation was studied in  \cite{Hartnoll:2008vx,Albash:2008eh}, where it was shown to lead to superconducting droplets that are exponentially confined near the origin (in our case the north pole). This behaviour is consistent with our findings for operators that are dual to tachyonic bulk scalars.
We conclude that in these cases, the rotating superconductor is equivalent to a static superconductor, immersed in a magnetic field of magnitude
\be
B_L =\frac{2m}{\sqrt{\Xi}e}\Omega_\infty\,,
\ee
where $\Omega_\infty$ is the angular momentum of the boundary theory and $m=|m_\Psi|$. Again, this field is weak, of order ${\cal O}\left( L^{-1} \right)$, in the limit we are considering. Furthermore, in the nonrelativistic regime $a^2 L^{-2}\ll 1$, it reduces to the relation between angular momentum and magnetic field first obtained by F. London \cite{London}, by considering his phenomenlogical theory for a rotating superconductor.
The work of \cite{Basu:2008st} studied the case where one adds linear momentum to the system. It would be interesting to see if their results can be seen in the present framework by zooming in on the equator of the solutions.

\section{Discussion}\setcounter{equation}{0}
In this paper we have demonstrated that finite-temperature gravitational backgrounds with non-zero angular momentum exhibit SSB much like their static limits. Furthermore we have seen that the additional dependence on rotation causes phenomena akin to those found in studies of superconductors in magnetic fields.

An important difference to this case is that there always exists a critical magnetic field $B_{c}$, even at very low temperature, above which the superconducting order is destroyed. We have seen that this is not always the case for the rotating superconductor on a sphere. This can be explained by noting that the local speed of rotation decreases as one moves away from the equator, and thus there can be regions of material, where the superconducting order is present, even when the equator moves at the speed of light. Near the poles we have seen that the field produced is precisely that predicted by F. London on the basis of his phenomenological theory of superconductors. The field is parametrically weak, of order ${\cal O}\left( L^{-1} \right)$, so  at low temperatures, it is not strong enough to destroy the superconducting order. This fact is also borne out in the full numerics, as exemplified in figure 5a.

 The London field is often written in terms of the dressed mass $m^*$ of the condensate, rather than the bare mass $m$. In our case, we find that $m^*=\frac{m}{\sqrt{\Xi}}$.  Thus the dressed mass corresponds to the relativistic mass of a particle of mass $m$ at the equator of the boundary sphere.

We have seen that even the lowest angular eigenfunction localizes the condensate in a droplet close to the equator for positive mass squared of the dual scalar field, and in a ring around the equator for negative mass squared. It is interesting that the localization behaviour of the condensate is qualitatively different for the case of positive bulk mass compared to negative mass (squared) and only in the latter case have we offered an explanation in terms of the material's London moment. It would clearly be an interesting question to understand what the different localization properties mean from the point of view of the boundary superconductor. It would appear that in one case the instablitity is towards forming a vortex anti-vortex pair localized on the antipodes, while in the other case we have a pair of lumps of superconducting material on the two poles.

Our analysis was done to first order in the small condensate. The localization behaviour observed in this paper would be expected in the early stages of vortex formation. At the level of approximation of this paper, the model does not allow for dynamically generated screening currents. However, \cite{Hartnoll:2008vx,Albash:2008eh} argue that going to higher order in the condensate will generate the screening currents necessary to confine the normal phase in vortex cores whose centres will allow magnetic flux to penetrate. Thus our results confirm the expectation that the rotating holographic superconductor has a London moment generating a London field.

It would be very interesting to extend the computations presented in this work to include the effects of backreaction of the scalar, both on the Maxwell field and on the gravitational background in order to understand the formation of vortices from the localized droplets in our system and to directly see the London field.

Independently of any holographic interpretation, the results of this paper indicate the existence of a new branch of stationary black hole solutions. Here we have found a marginal mode of the stationary asymptotically AdS black hole, which preserves axial symmetry. Just like in the static case, where the existence of a static marginal mode preserving the full spherical symmetry indicates the existence of a new branch of charged static solutions, here we expect the existence of a new branch of charged stationary solutions.  Clearly these last two issues are numerically more involved and will require more sophisticated methods than the ones discussed in this paper.

\begin{center}{\bf Acknowledgements}\end{center}

It is a pleasure to thank Eduard Antonyan, John McGreevy, Sean Hartnoll, Gustav Holzegel, David Tong, Arttu Rajantie and especially Toby Wiseman for helpful discussions and comments. I gratefully acknowledge generous support by EPSRC and Trinity College, Cambridge. The work presented here was initiated at MIT while I was supported by the NSF under grant PHY-0600465 and the DOE under collaborative research agreement DEFG02-05ER41360.

\appendix
\section{Separation of Variables}\setcounter{equation}{0}
While it can be shown by direct computation that the charged scalar field allows separable solutions, this procedure is as tedious as it is inelegant. We shall instead use the dyadic index formulation of Newman and Penrose \cite{Newman:1961qr} which is well suited to the present problem. A suitable Newman-Penrose tetrad is via the differential operators
\be\label{eq:npframe}
D=\ell^\mu\partial_\mu &=& \frac{1}{\Delta_r}\left( (r^2 + a^2)\partial_t  + \Delta_r\partial_r + a \Xi \partial_\phi \right)\nonumber\\
\Delta=n^\mu\partial_\mu &=& \frac{1}{2\rho^2}\left( (r^2 + a^2)\partial_t - \Delta_r\partial_r + a\Xi \partial_\phi  \right)\nonumber\\
\delta=m^\mu\partial_\mu &=& \frac{\sqrt{\Delta_\theta}}{\sqrt{2}\bar\rho}\left( \frac{ia\sin\theta}{\Delta_\theta}\partial_t + \partial_\theta + \frac{i \csc\theta}{\Delta_\theta}\Xi\partial_\phi \right)\nonumber\\
\bar\delta=\bar{m}^\mu\partial_\mu &=& \frac{\sqrt{\Delta_\theta}}{\sqrt{2}\bar\rho^*}\left(- \frac{ia\sin\theta}{\Delta_\theta}\partial_t + \partial_\theta - \frac{i \csc\theta}{\Delta_\theta}\Xi\partial_\phi \right)\,,
\ee
where all quantities are as defined in the main paper and in addition
\be
\bar\rho=r + ia \cos\theta
\ee
As ever, the metric takes the form
\be
\dd s^2 = 2 \left( m \otimes \bar{m} - \ell \otimes n \right)
\ee
in terms of the one-forms dual to \eqn{eq:npframe}. A simple representation of the non-vanishing spin coefficients is given by
\begin{align}
 \pi &=-\bar\delta \ln\bar\rho^*  \qquad\qquad\qquad\\
\rho &= D \ln\bar\rho^*\qquad\qquad\qquad\qquad~\,\tau = \delta\ln\bar\rho^*\,,\qquad\qquad\qquad\qquad\nonumber\\
\mu &=-\Delta \ln\bar\rho^*\qquad \qquad\qquad\quad~~\beta = -\frac{1}{4}\delta\ln\left(a^2\sin^2\theta\Delta_\theta  \right) \nonumber\\
\alpha &= \frac{1}{2}\bar\delta \ln\left(\frac{a\sin\theta\sqrt{\Delta_\theta}\Delta_r }{(\bar\rho^*)^2} \right)\,\quad \,\gamma= \frac{1}{2}\Delta\ln\left(\frac{a\sin\theta\sqrt{\Delta_\theta}\Delta_r }{(\bar\rho^*)^2} \right)\,,
\end{align}
where, in keeping with the notation introduced by Newman and Penrose, we also use the letter $\rho$ for one of the complex spin coefficients. Finally, the charged scalar field obeys the equation
\be
\eta^{(a)(b)}(\nabla_{(a)}+ieA_{(a)})(\nabla_{(b)}+ieA_{(b)})\Psi=m_\Psi^2\Psi
\ee
where we use Chandrasekhar's notation of putting indices that refer to the NP tetrad in brackets. 
Let us introduce the family of differential operators
\begin{align}
{\cal D}_n &= \partial_r - \frac{iK}{\Delta_r} + n \partial_r\ln\Delta_r\nonumber\\
{\cal D}^\dagger_n &=\partial_r + \frac{iK}{\Delta_r} + n \partial_r\ln\Delta_r\nonumber\\
{\cal L}^\dag_n &= \partial_\theta + \frac{H}{\Delta_\theta} + n\partial_\theta\ln\sin\theta\sqrt{\Delta_\theta}\nonumber\\
{\cal L}_n &= \partial_\theta - \frac{H}{\Delta_\theta} + n\partial_\theta\ln\sin\theta\sqrt{\Delta_\theta}
\end{align}
where acting on functions of the form
\be
\Psi(t,r,\theta,\phi)=e^{-i(\omega t + m \phi)}\Psi(r,\theta)\,,
\ee
the constants $K$ and $H$ are
\be
K=(r^2 + a^2)\omega + a m\Xi\,,\qquad \frac{H}{\Delta_\theta }=\left(\omega a \sin\theta + \frac{m\Xi}{\sin\theta}\right)\,.
\ee
It is then straightforward to show in a few steps that the charged scalar equation simplifies to the expression
\begin{align}
\left\{\frac{\Delta_r}{2\rho^2}\left[\left({\cal D}_1 - \frac{ieq_er}{\Delta_r}  \right)\left({\cal D}_0^\dag + \frac{eiq_er}{\Delta_r}  \right) + \left( {\cal D}_1^\dag + \frac{ieq_e r}{\Delta_r} \right) \left( {\cal D}_0 - \frac{eiq_e r}{\Delta_r} \right)  \right]\right.\nonumber\\
\left.+\frac{\sqrt{\Delta_\theta}}{2\rho^2}\left( {\cal L}_1^\dag\sqrt{\Delta_\theta}{\cal L}_0 + {\cal L}_1\sqrt{\Delta_\theta}{\cal L}_0^\dag     \right)\right\}\Phi=m_\Psi^2\Phi
\end{align}
Now, notice that the operators ${\cal D}_{0,1}$ and their hermitian conjugates are purely radial, while the operators ${\cal L}_{0,1}$ and their hermitian conjugates are purely angular. Thus the equation is separable upon multiplying by $\rho^2$, resulting in the two equations
\be
\left[\Delta_r (\cD_1 \cD_0^\dag +\cD_1^\dag \cD_0) + 2ie q_e r (\cD_0 - \cD_0^\dag) + 2\frac{e^2 q_e^2 r^2}{\Delta_r} - 2(r^2 +a^2)m_\Psi^2-2\lambda \right]R(r) = 0
\ee
and
\be
\left[ \left(\sqrt{\Delta_\theta} \cL_1 \sqrt{\Delta_\theta}\cL_0^\dag  + \sqrt{\Delta_\theta}\cL_1^\dag\sqrt{\Delta_\theta} \cL_0\right) +2a^2\sin^2\theta m_\Psi^2+2\lambda\right] \Theta(\theta)=0\,.
\ee
Substituting the definitions of the differential operators ${\cal L}_{0,1}$ and ${\cal D}_{0,1}$ results in the radial and angular equations quoted in the main text modulo a gauge choice, on which we now elaborate.
\section{Details on Numerical Methods}\setcounter{equation}{0}
This appendix contains details on the gauge choice used in equation \eqn{eq:radialI} and gives details on the numerical algorithms used to compute the separation constants used in the bulk of the paper.
\subsection{Gauge Choice and Regular Solutions}
The potential contribution to the radial equation contains an effective mass-squared term proportional to the square of the gauge field. This term seemingly blows up at the horizon unless the gauge field $A$ goes to zero there. Notice however, that the function $R(r)$ that we are solving for is itself not a gauge invariant quantity. In particular, it has a gauge-dependent phase. Thus if we choose the radial function to be
\be
e^{ie\varphi(r)}R(r)\,,
\ee
we can remove the apparent singular behaviour of the potential at the horizon with a particular gauge choice for $\varphi(r)$. A computation shows that the function
\be
\varphi(r) = \frac{q_e}{r+}\int\frac{\dd r}{h(r)} = \frac{q_er_*}{r_+}\,,
\ee
which we recognize as the tortoise coordinate \eqn{eq:tortoise}, removes the apparent singular behaviour at the horizon. With this choice of phase, the gauge-invariant part of the radial function satisfies the equation \eqn{eq:radialI}.

\subsection{Angular Eigenvalues and Eigenfunctions}
The separated equations are coupled through the separation constant $\lambda$. It is thus important to have precise numerical methods to determine it. With $x=\cos\theta$ equation \eqn{eq:angularI} reads
\be\label{eq:kerradsangular}
\sqrt{1-\alpha^2 x^2}\frac{d}{dx}\left[\sqrt{1-\alpha^2 x^2}(1-x^2)\frac{dS_{\ell m}^\gamma}{dx}  \right] - \alpha^2 x (1-x^2)\frac{dS_{\ell m}^\gamma}{dx} & &\nonumber\\+\left[\lambda + a^2 m^2_\Psi(1- x^2)- \frac{m^2(1-\alpha^2)}{1-x^2} \right]S_{\ell m}^\gamma(x)&=&0\,,
\ee
where we have briefly reinstated the azimuthal quantum number $m$. In the limit $L\rightarrow\infty$, equation \eqn{eq:kerradsangular} goes over into the spheroidal equation and the eigenfunctions $S_{\ell m}^\gamma(x)$ are the spheroidal harmonics.  In many references on spheroidal harmonics the parameter $a^2 m^2_\Psi$ is called $\gamma^2$.  In the limit $a\rightarrow 0$ it becomes the associated Legendre equation.

In this paper, we only consider $S_{\ell m}^{\gamma \pm}(x,\lambda)$ for $m=0$, which is all we need in determining the critical temperature. Note, however, that the angular eigenvalue of the (AdS) spheroidal equation depends on $m$, unlike the more well-known case of the spherical harmonics (i.e. the Legendre polynomials). However, our procedures work equally well for non-zero values of $m$.
\subsubsection{Numerics}
We now describe the numerical procedure used to extract the eigenvalues $\lambda$ from this equation. It is evident that solutions of equation \eqn{eq:kerradsangular} are either even or odd functions. We will make use of this crucial fact, which makes it easier to use a simple shooting method:
\begin{enumerate}
\item Obtain series expansions of $S^\gamma_{\ell m}$ at $x=\pm 1$. There is a regular branch and a logarithmic branch of solutions. As usual we select the regular branch. We set the following boundary conditions
\be
S_{\ell m}^\gamma(-1) = S_{\ell m}^\gamma(1)=1 \qquad S_{\ell m}^\gamma(x)\quad{\rm is}\,\,{\rm even}\nonumber\\
 S_{\ell m}^\gamma(-1) = -S_{\ell m}^\gamma(1)=1 \qquad S_{\ell m}^\gamma(x)\quad{\rm is}\,\,{\rm odd}
\ee
The choice of the magnitude of the eigenfunctions at either end of the interval is clearly just a choice of normalization.
\item Numerically integrate inwards to the origin from both ends for a fine grid of values of the eigenvalue $\lambda$ in order to match the Frobenius expansions at either end of the intervals. This gives rise to a series of functions $S_{\ell m}^{\gamma \pm}(x,\lambda)$, where the $\pm$ refers to the numerical integration from $\pm 1$.
\item For odd functions the first derivatives of $S_{\ell m}^{\gamma \pm}(x,\lambda)$ are guaranteed to coincide at the origin. Thus we define the function
$$
O(\lambda) = S_{\ell m}^{\gamma, +}(0,\lambda) - S_{\ell m}^{\gamma, -}(0,\lambda)
$$
by interpolating the values of the discrete grid of values of $\lambda$. For even eigenfunctions, we are guaranteed that the value of $S_{\ell m}^{\gamma \pm}(x,\lambda)$ matches at the origin. Thus we define
$$
E(\lambda) = \left(S_{\ell m}^{\gamma, +}\right)'(0,\lambda) - \left(S_{\ell m}^{\gamma, -}\right)'(0,\lambda)
$$
\item Find the zeroes of $E(\lambda)$ and $O(\lambda)$. Every root corresponds to an eigenvalue of the angular equation.
\end{enumerate}
We now tabulate (a subset of) the $\lambda$ values that were used to find the critical temperatures displayed in figure \ref{fig:phasediagram}.
Table 1 lists a representative sample of eigenvalues to five significant digits that were used in obtaining the phase curves of critical temperature versus rotation.    Our numerical algorithms in fact allow a far higher accuracy, but it would be impractical and of little value to list more significant digits here.

\begin{table}[h!]
\caption{{\small Angular eigenvalues of AdS spheroidal harmonics. The first line gives the deformation parameter $\alpha=a L^{-1}$ and the following two lines supply the angular eigenvalues for the two mass-squared values used to compute the phase diagram.}}
\begin{center}
{\footnotesize
\begin{tabular}{|c|c|c|c|c|c|c|c|}
\hline
$\alpha$ & 0 &0.1& 0.2  &0.4 & 0.6 & 0.8 & 0.9 \\
\hline\hline
$m^2_\Psi L^2=-2$&0 &0.0133& 0.0532 &0.2115 &0.4690 & 0.8086 & 0.9921 \\
\hline
$m^2_\Psi L^2=4$ &0 &-0.0266&-0.1071&-0.4340 & -0.9997 & -1.8393 & -2.3764\\
\hline
\end{tabular}}
\end{center}
\end{table}

\end{document}